\documentstyle[12pt]{article}
\begin{document}
\baselineskip=14truept plus 0.2truept minus 0.2truept

\begin{center}

{\bf On some singularities of the correlation functions that determine neutrino opacities }

\end{center}

\medskip
\begin{center}

R. F. Sawyer

\end{center}

\begin{center}

Department of Physics, 
University of California at Santa Barbara
\\
Santa Barbara, California 93106

\end{center}

\medskip

\begin{abstract}
Certain perturbation graphs in the calculation of the effects of the medium on neutrino scattering in supernova matter have a nonintegrable singularity in a physical region. A number of papers have addressed the apparent pathology through an ansatz that invokes higher order (rescattering) effects. Taking the Gamow-Teller terms as an example, we display an expression for the spin-spin correlation  function that determines the cross-sections. It is clear from the form that there are no pathologies in the order by order perturbation expansion. Explicit formulae are given for a simple case, leading to an answer that is very different from one given by other authors.
 \end{abstract}

There is a considerable literature$^{1-6}$ directed at the time dependent spin fluctuations in the hot nuclear matter of the postshock supernova environment, with application to the neutrino physics in the medium. The nucleon spin correlation functions are of particular importance in these problems because the Gamow-Teller part of the weak interaction supplies roughly 3/4 of the rate for neutrino-nucleon processes. The spin (and spin-isospin) correlation functions determine the GT reaction rates and are known to be strongly affected by the N-N interactions in the medium$^{7-11}$. However, there has been disagreement both about the relative importance of the different corrections, and about the detailed mechanics of calculation. Much of the work referred to above attempts to go beyond perturbation theory; but one of the disagreements about method is rooted in the properties of perturbation theory. This has to do with the nonintegrable perturbative singularities in the time Fourier transform variable, $\omega$, coming from simple graphs in an approach based on T matrices. The authors of refs  \cite{RS}-\cite{RSold} advocate the regulation of these singularities by rescattering affects that they estimate through a physical rather than an analytical argument. In contrast, I have argued that these singularities are regulated, in the calculation of  neutrino scattering effects, by contributions from ``medium dependent wave function renormalization"$^{12}$.

In the present note we consider the Fourier transform of the time-dependent spin correlation function in an interacting system, and present a formal expression that is well adapted for the demonstration that the singularities must disappear, order by order, in physical results. The end results are the same as those provided by a T matrix approach followed by regularization via medium-dependent wave function renormalization, but the latter procedure is no longer required. Our procedure is also simpler, both conceptually and in the application at hand, than is the construction in the imaginary time formalism of the requisite thermal Green's functions, followed by the continuations needed to get the real time correlation function. When the spin densities are replaced by the complete set of weak current operators, the approach laid out below also is well adapted to perturbative consideration of the correlation functions for all of the currents. It should be especially useful in resolving the disagreements that persist in the calculation of 
medium dependent wave function renormalization that has been needed in calculating the effects of the photon bath on the weak interaction rates in the early universe, just before nucleosynthesis$^{13-16}$. 

The Fourier transform of the correlation functions for the spin densities of an N particle system is given by,
\begin{equation}
S_{i,j}(q,\omega)=Z^{-1}\int d^3x dt \;e^{-i \bf q\cdot
  x}\rm\it e^{i\omega t}Tr\bigl[e^{-\beta H}\sigma_i(\bf x \rm\it , t)\sigma_j(0,0)\bigr],
\label{1}
\end{equation}
where the $\sigma_i(\bf x \rm\it , t)$  is the spin density operator in the Heisenberg picture, and $Z$ is the partition function.
  
To determine the rates of neutrino reactions in the medium this correlation function is integrated with respect to a weight function, $g_{i,j}(q,\omega)$,
\begin{equation}
W=\sum_{i,j}\int d^3q d\omega S_{i,j}(\bf q \rm,\omega)g_{i,j}(\bf q \rm,\omega),
\label{2}
\end{equation}
where $(\bf q \rm,\omega)$ is the momentum-energy transfer to the leptons and where the weight function is determined  kinematically, depending on the reaction under consideration. We need to distinguish two cases:
\begin{enumerate} 
\item The support of the function, $g(q,\omega)$ includes a region around $\omega=0$; 
\item The function $g(q,\omega)$ vanishes for $\omega<0$ or for $\omega>0$.
\end{enumerate}
Neutral current $\nu$ scattering is an example of case \#1 and 
$\nu$ pair production or absorption are examples of case \#2.  

To extract terms in the perturbation development of (\ref{1}) we make a transformation to an interaction picture based at time $t=0$,
\begin{equation}
\sigma_i(\bf x \rm \it,t)=U(0,t)\sigma^{(I)}_i(\bf x \rm \it,t)U(t,0),
\label{3}
\end{equation}
 where $U(t_1,t_2)$ is the interaction picture time translation operator,
\begin{eqnarray}
 U(t_2,t_1)=T[\exp(-i\int_{t_1}^{t_2} dt H_I(t))]= 
\nonumber\\ 
1-i \int_{t_1}^{t_2} dt' H_I(t')-\int_{t_1}^{t_2}dt'H_I(t')\int_{t_1}^{t'}dt''H_I(t'')+.....
\label{4}
\end{eqnarray}
$H_I(t)$ is the interaction Hamiltonian in the interaction picture. In (\ref{4}) we write the familiar perturbation expansion. It defines a continuation for complex $(t_1,t_2)$, in the case in which $H_I(t)$ is an analytic function of $t$, so that the integrals are independent of path. With the continuation so defined, we also have the result,
\begin{equation}
e^{-\beta H}=  e^{-\beta H_0}U(-i\beta,0),
\label{5}
\end{equation}
which can be easily verified by observing that the continuation of (\ref{4}) provides the "$\tau$-ordered" operator from finite- temperature equilibrium quantum mechanics$^{17}$,
\begin{equation}
U(-i\beta,0)=T[\exp( -\int_0^{\beta}d\tau H_I(-i\tau))].
\label{6}
\end{equation}
The interaction picture spin-density operator, $\sigma^I_i(\bf x \rm,t)$ (which is the axial vector current for nonrelativistic nucleons), can be written in terms of the Pauli matrices for the individual particles, $\sigma_i^{(\alpha)}$
\begin{equation}
 \sigma^{I}_i(\bf x \rm\it, t)=\sum_{\alpha=1}^N\delta(\bf x-x_{\rm \it \alpha}-\rm\it m^{-1} 
 \bf p_{\alpha}\rm\it t+ \bf c_{\rm\it \alpha}\rm\it) \sigma_i^{(\alpha)},
\label{8}
\end{equation}
where $\bf x_{\rm\it \alpha}$ is the Schr\"odinger position operator and $\bf p_{\rm \it \alpha}$ the momentum operator for the $\alpha$'th particle and the $\bf c_{\rm\it \alpha}$'s are irrelevant constants.
We can use (\ref{3}) and (\ref{5}) to write (1) as
\begin{equation}
S_{i,j}(q,\omega)=Z^{-1}\int dx dt \;e^{-i \bf q \cdot 
  x \rm\it}e^{i\omega t}Tr\Bigl[e^{-\beta H_0}U(-i\beta,t)\sigma^{(I)}_i(\bf x \rm\it,t)U(t,0) \sigma^{(I)}_j (0,0)\Bigr],
\label{7}
\end{equation}
where we have used the composition properties of the $U$ operator, $U(-i\beta,0)U(0,t)=U(-i\beta,t)$.

Eq. (\ref{7}) is our basic result. The interaction Hamiltonian enters only through the $U $ operators, and it remains to expand them in the power series (\ref{4}) and take the matrix elements. We note some features before giving an example.

A) Having based the interaction picture at t=0, we will generate more terms in a given  perturbation theory time integral than one encounters in constructing the correlation function from on-shell T matrix elements weighted by statistical factors. This is because the T matrix elements are calculated in an interaction picture based at $t=-\infty$, with adiabatic cutting off, and in a typical time integration one therefore gets a contribution only from the upper limit, whereas in our time integrals in the expansion of matrix elements of $U$ there are finite contributions from both limits. The cost of the T matrix approach, is considerable, however, in requirinq a ``medium dependent wave-function renormalization"$^{12}$, that must be done carefully. The need for this procedure is a consequence of turning the interaction off asymptotically when the interacting particles do not become separated at long times. This is analogous to the case of a field theory in a vacuum, where the self interactions, which do not disappear at large times, lead to a wave function renormalizations for all participating particles, with the difference that now it is the interactions with the medium and not the self-interactions that must be taken into account.

As remarked above, if one follows a T matrix approach, but without including this wave function renormalization , and calculates $W$, for case \#1, one encounters non-integrable singularities in the lowest- order perturbation correction to the free-particle correlation function. In contrast, the basing of the interaction picture at t=0, as in (\ref{7}) avoids both any singular terms in the straightforward order by order expansion of $W$, and the conceptual complications of the T matrix approach as well.

B) The operator $U(-i\beta,t)$ in (\ref{7}) combines part of what is in T matrix language the "scattering" correction (the $t$ to $0$ part) with the correction to the statistical factor (the $0$ to $-i\beta$ part). We can now do the integrations for the combined factor directly, saving in the number of terms generated. Thus it seems best on all counts to use the form,(\ref{7}), with the interaction picture based at $t=0$, as the basis for calculating correlation functions perturbatively. 

As an illustration, we take the model of Raffelt and Sigl$^1$, where only the auto-correlation function of a single particle that moves in one dimension is considered and where the interaction is through a fixed spin-dependent potential. In ref. \cite{RS}, this potential, designed to simulate the rest of the medium, is one produced by a set classical moments of random direction, situated at random points, $x_\alpha$, each with an (interaction picture) interaction with the particle,

\begin{equation}
<s,p|H_I^{(\alpha)}(t)|s',p'>=v(p-p')\stackrel{\rightarrow}{\mu_\alpha}\cdot \stackrel{\rightarrow}{\sigma}_{ s,s'}e^{i  (p-p')\cdot x_{\alpha} }e^{i(E_p-E_{p'})t},
\label{9}
\end{equation}
where $\mu$ is the classical moment and the states of the particle are labelled by a single momentum and spin, $p,s$, and $E,E'$ are the energies of the states labelled by $p,p'$. For our demonstration, it suffices to take a single one of these sources. At the end of the calculation the correlation function will get averaged over all positions of this source, so that the implicit use of translation invariance in (\ref{1}), in using the points, $x$ and $0$, is allowed. In ref.\cite{RS} the authors find that their results are quite independent of the momentum transfer, $q$, over the range of $q$ defined by the weight function, $g_{i,j}(q,\omega)$. For our tests we set $q=0$ in the correlation function; the formulae below would be slightly more complex if we kept the $q$ dependence, and the conclusions the same.  Then, using (\ref{8}) with N=1, we perform the $x$ integral in (\ref{7}) obtaining for the second order contribution,
\begin{equation}
S_{i,j}(q=0,\omega)=Z^{-1}\int dt \; e^{i\omega t}Tr\Bigl[e^{-\beta  H_0}U(-i\beta,t)\sigma_i U(t,0) \sigma_j \Bigr],
\label{10}
\end{equation}
where $\sigma_i$ is simply the Pauli matrix for the single dynamical particle.
To evaluate (\ref{10}) to second order in the strength parameter, $\mu$, we write the expansion (\ref{4}) as $U=U^{(0)}+U^{(1)}+U^{(2)}$ and perform the time integrals, obtaining,
\newpage
\begin{eqnarray}
W=Z^{-1}(2\pi)^{-1}\sum_{i,j}\sum_{s=1,2} \int dq dp d\omega g_{i,j}(q,\omega) \int dt e^{i\omega t}<p,s|e^{-\beta H_0 }\Bigl[\sigma_i U^{(2)}(t,0)\sigma _j +
\nonumber\\
\,
\nonumber\\
 U^{(2)}(-i\beta,t)\sigma_i \sigma_j+U^{(1)}(-i\beta,t)\sigma_i U^{(1)}(t,0)\sigma_j \Bigr]|p,s>\;\;\; . \;\;\;\;\;
\label{11}
\end{eqnarray}
Using the interaction Hamiltonian (\ref{9}) in the terms of (\ref{4}) and performing the time integrals we obtain
\begin{eqnarray}
&W=Z^{-1}\pi^{-1}\mu^2 \int d\omega g_1(\omega)\int \frac{dp dp_1e^{-\beta E_1}}{  |v(p_1-p)|^2}
\Bigl[\frac{\frac{16}{3}\delta (\omega-E_1+E)+[2\beta (E_1-E)-\frac{16}{3}]\delta(\omega)}{(E_1-E)^2}\Bigr]
\nonumber\\
&\,
\nonumber\\
&=Z^{-1}\pi^{-1}\mu^2\int \frac{dp dp_1e^{-\beta E_1 }}{|
v(p_1-p)|^2} \Bigl[\frac{\frac{16}{3}g_1(E_1-E)+[2\beta (E_1-E)-\frac{16}{3}]g_1(0)}{(E_1-E)^2}\Bigr]\,
\label{12}
\end{eqnarray}
where $g_1(\omega)=\sum_{i,j}\int dq g_{i,j}(q,\omega)$, and where we have averaged over the directions of $\stackrel{\rightarrow}{\mu}$. In  case \#1, for which the prototype is neutrino scattering, we can now see from (\ref{12})  both how it may appear that there is singular behavior that needs regulation, and why that is not the case. If in the middle step of (\ref{12}) we do the $p_1$ integration before the $\omega$ integration, and take just the first term in the square bracket, we do indeed get a $\omega^{-2}$ singularity, and it is not removed by any other $\omega^{-2}$ singularity. This first term is the whole of the ``perturbation" result quoted in ref.\cite{RS} and essentially that of similar expressions in refs. \cite{blah1}-\cite{RSold}. But if we look at the results of integrating over $\omega$ first, in the last line of (\ref{12}), where the singularity of this first term turns into $(E-E_1)^{-2}$, we see how the other terms remove both the second and first order singularities, leaving a regular remainder. The physics is simple; in the middle step of (\ref{12}) the $\delta(\omega-E_1+E)$ term has to do with the medium being left excited by the transfer of energy $\omega$, whereas the cancelling term with $\delta (\omega)$ is from the process in which the medium is left unexcited. This is the term that is introduced by the ``medium dependent wave function renormalization" in ref.\cite{saw2}. We see that in the present more systematic approach it appears from a completely mechanical calculation of the correlation function.  Thus we conclude that the replacement by a  Lorentzian form,  $\omega^{-2} \rightarrow (\omega^2+\Lambda^2)^{-1}$, as in refs.   \cite{RS} -\cite{RSold}, is neither necessary nor justified.

Turning to case \#2, for which the prototype is neutrino pair production, we observe that, to make sense, the weighting function $g(\omega)$ must vanish at $\omega=0$; in the neutrino pair production problem it vanishes as $\omega^3$. The $\delta(\omega)$ term  contributes nothing to this process and the $\omega^{-2}$ singularity is harmless. Therefore there is again no reason to regulate the $\omega^{-2}$ terms in the correlation function $\omega^{-2}$. 

To make this point more sharply it may be useful to redescribe the problem that is being addressed here. We started with the question of how the interactions of nucleons in the medium affect the neutrino scattering rate. Suppose we ask instead how the interaction with a neutrino would affect the scattering of a nucleon within the matter. Our amplitudes would have a $\omega^{-1}$ singularity, and the rate corrections an $\omega^{-2}$ singularity, the latter again being removed, in an inclusive calculation, by a non-analytic term, $\delta (\omega)$. Notwithstanding this order-by-order removal in the inclusive rate, which has to do with the interference of connected and disconnected T matrix graphs in the construction of the correlation function, let's look at the analytic (and totally connected) part of the amplitude by itself. The singularity comes from the graphs in which the weak current, which transmits energy $\omega$, attaches to an external nucleon line in the nuclear process. It is a simple pole singularity in the amplitude, no matter how complex an N-N amplitude it attaches to, and the same holds true if we consider many-nucleon amplitudes. The point is that in every order in the nucleon-nucleon interaction, the singularity is still a simple pole. If the sum of graphs were to move the position of the singularity, as in the case of propagator modifications, or to exponentiate the singularity, as in the infrared problem, we would have found higher order singularites in the higher order graphs. We conclude again that the regulation proposed in refs. \cite{RS}-\cite{RSold} does not occur. This is not to say that multinucleon effects cannot produce interesting modifications of the exact correlation functions for small $\omega$. However, we are skeptical that the mock-up of many body effects used in ref.\cite{RS}, in which all but one nucleon is replaced by a set of fixed potentials with randomized positions for the sources, can be illuminating in discussing this behavior.

A numerical nonperturbative calculation is done in ref.\cite{RS} as a function of $\omega$, but only for $\omega\ne 0$. The numerical calculation cannot be extended to $\omega=0$ to produce the $\delta(\omega)$ terms that are required to cancel the singularity in case \#1, so that the calculation itself does not address the issues raised in the present paper. However, we can compare with our result (\ref{12}) the result, quoted in \cite{RS}, for the regulated perturbation result in the model. We do this for case \#1, taking neutrino scattering kinematics, and in one dimension. The ansatz of ref's  \cite{RS}-\cite{RSold}leads to the replacement of (\ref{12})  by
\begin{equation}
W^{reg}=Z^{-1}\pi^{-1}\mu^2\int \frac{dp dp_1e^{-\beta E_1 }}{v(p_1-p)|^2 }|
\Bigl[\frac{\frac{16}{3}g_1(E_1-E)}{(E_1-E)^2+\Gamma^2}\Bigr].\;\;
\label{13}
\end{equation}
Using parameters from ref.\cite{RS} we take $v(p)=v_0 \exp[-(bp)^2/4]$, where $b$=1F, and $\Gamma=.5(T/1 \rm MeV)^{1/2}MeV$. For $g_1(\omega)$ we take the function that comes from the integrals over the initial and final neutrino phase spaces, weighted with the appropriate Fermi factors and energy conserving delta function, of the leptonic part of the squared Gamow-Teller matrix element. We do this in a one-dimensional world, for consistency. Since we are interested only in comparing (\ref{12}) and (\ref{13}), we can normalize to $g_1(0)=1$. We obtain,
\begin{eqnarray}
g_1(\omega)=2 \theta (\omega)(1-e^{-\beta \omega})^{-1}[\log (2)-\log (1+e^{-\beta \omega})]+
\nonumber\\
+2 \theta (-\omega)(1-e^{-\beta \omega})^{-1}[-\log (2)+\log (1+e^{\beta \omega})].
\end{eqnarray}

We have calculated the ratio of the rate corrections given in (\ref{12}) and (\ref{13}) for a range of temperatures from  2.5 MeV-100 MeV. $W$ is negative and $W^{reg}$ positive for all temperatures, with the ratio $W/W^{reg}$ ranging from $-$.70 at $T$=2.5 MeV to $-$.15 at $T$=100 MeV. Note that these correction terms are proportional to the same unspecified strength parameter. The perturbation theory is justified, at best, only when this strength is small enough so that the perturbation is smaller than the unperturbed rate; we are not predicting negative rates.

A similar reduction in rate coming from interactions with the surroundings has been found in a related three-dimensional model$^{12}$. Reverting to a T matrix description, the second order contribution has two parts. The first reflects a change in state of the medium and is positive, since it is, loosely speaking, connected to the opening of another channel. The second reflects a second order correction to the wave-function of the nucleons in the medium, and leaves the medium unexcited, so that there can be an interference between zeroth and second order amplitudes. This term is negative and outweighs the first term. 

These comments do not apply to the neutrino pair production rate; in this case there is no zeroth order term, the leading second order term is positive, and the discrepancy between the results of our method and the results of refs.\cite{RS}-\cite{RSold}is simply the change introduced by the presence, in the latter, of the arbitrary regulator, $\Lambda$.

Finally, we note that the model defined by the single particle interacting with classical magnetic moments, with form factors, does not simulate important, and calculated, effects of true many-body interactions on current correlation functions. The ring graphs, which  involve at least two dynamical particles, are not included in any sense in the calculations of ref.\cite{RS}. The need for the ring sum is not an arbitrary singling out of a special set of graphs; as laid out in ref.\cite{BS}, among other things, the ring sum embodies the long-wavelength limits of the four correlation functions that determine the neutrino scattering process. The ring corrections lead to enhancement or suppression, depending on the sign of the force, if we are doing actual perturbation theory with a potential; or depending on the signs of Fermi liquid parameters if we follow the approach of \cite{BS}. This is in contrast to the perturbative terms produced in the above model, which do not depend on the sign of the interaction.

Thanks are due to Adam Burrows for a number of discussions and to Lowell Brown for instruction in thermal quantum mechanics.

\end{document}